\documentclass[altaffilletter,superscriptaddress,amsmath,amssymb,twocolumn]{revtex4-1}
\usepackage{graphicx}
\usepackage{textcomp}

\begin{document}

\title{Broadband monolithic extractor for metal-metal waveguide based terahertz quantum cascade laser frequency combs}%

\author{Markus R\"osch}%
\email{mroesch@phys.ethz.ch}
\author{Ileana-Cristina Benea-Chelmus}
\author{Christopher Bonzon}
\author{Martin J. S\"uess}
\author{Mattias Beck}
\author{J\'er\^ome Faist}
\author{Giacomo Scalari}
\email{scalari@phys.ethz.ch}
\affiliation{ETH Zurich, Institute of Quantum Electronics, Auguste-Piccard-Hof 1, Zurich 8093, Switzerland}

\begin{abstract}
 We present a monolithic solution to extract efficiently light from terahertz quantum cascade lasers with metal-metal waveguides suitable for broadband frequency comb applications. The design is optimized for a bandwidth of 500\,GHz around a center frequency of 2.6\,THz. A five-fold increase in total output power is observed compared to standard metal-metal waveguides. The extractor features a single-lobed far-field pattern and increases the frequency comb dynamical range to cover more than 50\,\% of the laser dynamic range. Frequency comb operation up to a spectral bandwidth of 670\,GHz is achieved.
\end{abstract}

\maketitle

\begin{figure}[tbh]
  \centering
  \includegraphics[scale=1]{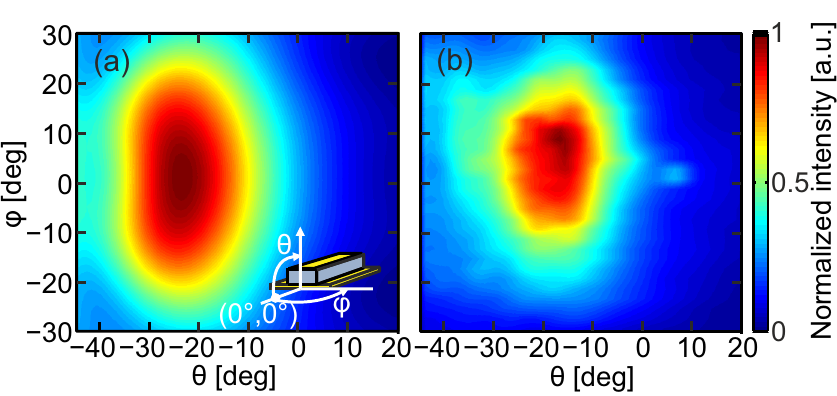}
\caption{Far-field of the extractor: (a) Simulated far-field pattern of the extractor simulated with CST Microwave studio$^{\text{\textregistered}}$. The inset displays the orientation of the laser with respect to the detector. (b) Measured far-field of a 2.5\,mm long and 60\,\textmu m wide laser with an extractor consisting of 5 patches in quasi-CW operation (95\% duty cycle) at a temperature of 20 Kelvin for a driving current of 405\,mA.}
\label{fig:farfield}
\end{figure}

\begin{figure}[tbh]
  \centering
  \includegraphics[scale=1]{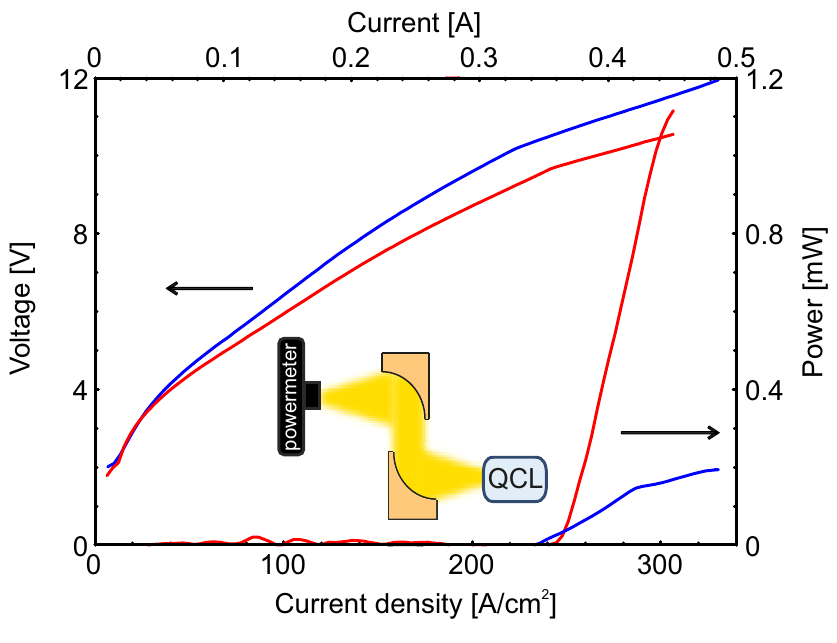}
\caption{Power measurements: Light-current-voltage characteristics of a 2.5\,mm long and 60\,\textmu m wide laser with an extractor consisting of 5 patches in CW operation (red) compared to a laser with almost identical active area (2\,mm long and 70\,\textmu m wide) without extractor. The inset shows schematically the setup for the power measurement using two off-axis parabolic mirrors ($f/3$). A calibrated absolute powermeter (Ophir: 3A-P-THz) has been used to measure the QCL power.}
\label{fig:liv}
\end{figure}

\begin{figure}[t]
  \centering
  \includegraphics[scale=1]{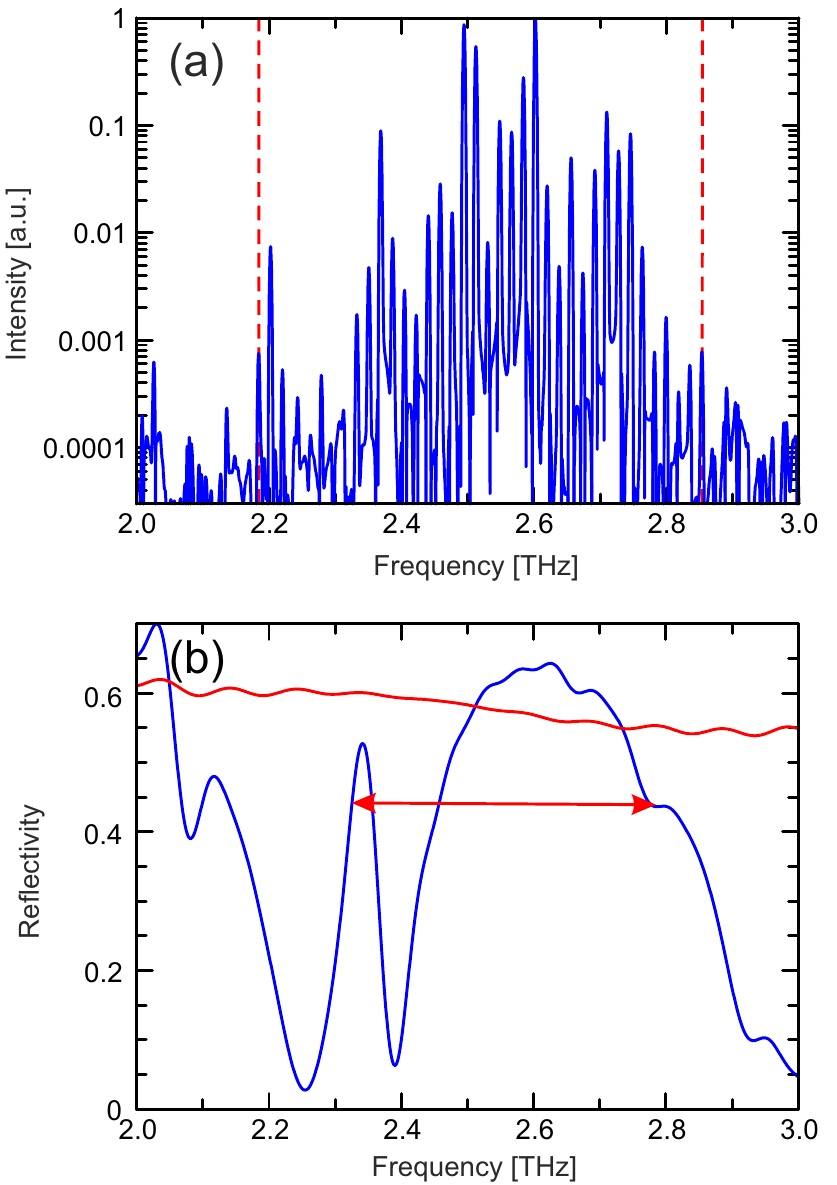}
\caption{Optical spectrum: (a) A 2.5\,mm long and 60\,\textmu m wide laser with an extractor consisting of 5 patches in CW operation was spectrally measured using an under-vacuum FTIR. The laser was operated at 420\,mA bias at a temperature of 18 Kelvin. Only lasing on fundamental lateral modes can be observed. The first and last continuous lateral mode are marked by the red dashed lines. (b) Reflectivity for a standard metal-metal waveguide facet (red), and for the reported extractor (blue) simulated with CST Microwave Studio$^{\text{\textregistered}}$. The red arrow indicates the 30\,dB limit of the optical spectrum from (a).}
\label{fig:spectra}
\end{figure}

\begin{figure}[t]
  \centering
  \includegraphics[scale=1]{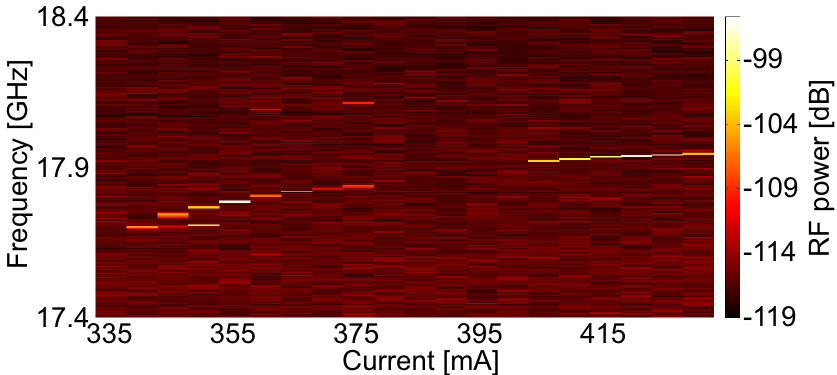}
\caption{Beatnote map: The frequency of the intermode beatnote of a 2.5\,mm long and 60\,\textmu m wide laser with an extractor consisting of 5 patches is measured as a function of driving current in CW operation  at a temperature of 20 Kelvin showing comb operation over most of the dynamical range of the laser.}
\label{fig:beatnote}
\end{figure}

Recent years have brought up a new family of frequency combs (FC) in the mid-infrared and terahertz (THz) frequency ranges formed by four-wave mixing (FWM) in quantum cascade lasers (QCLs)  \cite{Hugi2012,Burghoff2014,Roesch2014,faist2016combs}. Due to their internal locking mechanism, these devices are very compact and therefore very interesting for integrated spectroscopy systems \cite{villares2014}. For spectroscopic applications a broad spectrum is desired. Recent work on THz QCLs has shown the possibility of achieving octave-spanning lasers \cite{Roesch2014}. Such lasers would be ideal candidates for octave-spanning, fully stabilized FCs \cite{telle1999f2f,jones2000f2f,Udem2002}. \\
All broadband THz QCLs, including THz QCL FCs, are based on a metal-metal waveguide \cite{turcinkova2011,Burghoff2014,Roesch2014}. The metal-metal waveguide is a cutoff-free waveguide with very flat losses and low dispersion over the entire THz frequency range ideal for FC operation \cite{williams2003,Roesch2014,bachmann2016setback,bachmann2016dispersion}. The major drawback of the metal-metal waveguide is its facets with sub-wavelength dimensions. This results in a high reflectivity and a strongly diffracted far-field \cite{williams2003,kohen2005,Adam2006}. For most applications, including FC based spectroscopy, a high power and a well-defined far-field are essential. For THz QCLs with metal-metal waveguides several attempts have been undertaken to overcome the limitations of the subwavelength facet. A very successful approach uses a distributed feedback grating (DFB) matching the 3rd order Bragg condition \cite{amanti2009}. Their almost Gaussian beam features high power in single mode operation. Similar results were achieved using various one dimensional (1D) or two dimensional (2D) confinement such as 1D photonic heterostructures \cite{xu2012,xu2014}, 2D annular DFBs \cite{liang2013}, 2D photonic crystal lasers \cite{chassagneux2009,chassagneux2010,halioua2014}, photonic quasi-crystals \cite{vitiello2014}, metasurface external cavity lasers \cite{xu2015qclvecsel}, integrated patch antennae \cite{bonzon2014patch,bosco2016,justen2016}, plasmonic lasers \cite{wu2016}, and random lasers \cite{schoenhuber2016}. However, all these techniques are based on frequency selection and yield single mode operation.\\
Some other techniques have been reported to improve the far-field performance of multi-mode THz QCLs. One can either place a silicon lens in front of a standard ridge \cite{lee2007lens}, or a horn antenna is post-processed to a laser \cite{amanti2007}. Both approaches have the drawback of being non-monolithic and requiring advanced post-processing steps. A simpler approach has recently been reported, which uses a post-process focused ion beam (FIB) step to remove part of the top metallization to form a horn antenna \cite{brewer2014,wang2016}. However, the far-fields are not fully single lobed. In addition, this approach has not been investigated concerning its complience with broadband FC operation. \\
We report here a technique to improve far-field and power performance in a multi-mode regime which can be implemented during the standard processing steps. An extractor structure has been implemented on one side of the laser. This helps to increase the output power and leads to a better confined far-field while keeping the broadband frequency range. The reported extractors are also compatible with FC operation. \\
The extractor structure we report on here, is conceptionally an end-fire antenna. The antenna is integrated in the waveguide of the QCL similar to the patch antennae reported in Ref. \cite{bonzon2014patch,bosco2016,justen2016}. To select a forward directioned output beam the starting point for the design is the 3rd order Bragg condition for the central wavelength of 2.6\,THz. In contrast to a 3rd order DFB, not the entire waveguide is patterned. Instead only 5 periods of the grating were used, leaving the major part of the waveguide unpatterned. The grating will not be used for the optical feedback of the laser cavity, as in a DFB, but only to shape the far-field of the extracted light. \\
\begin{figure}[tbh]
  \centering
  \includegraphics[scale=1]{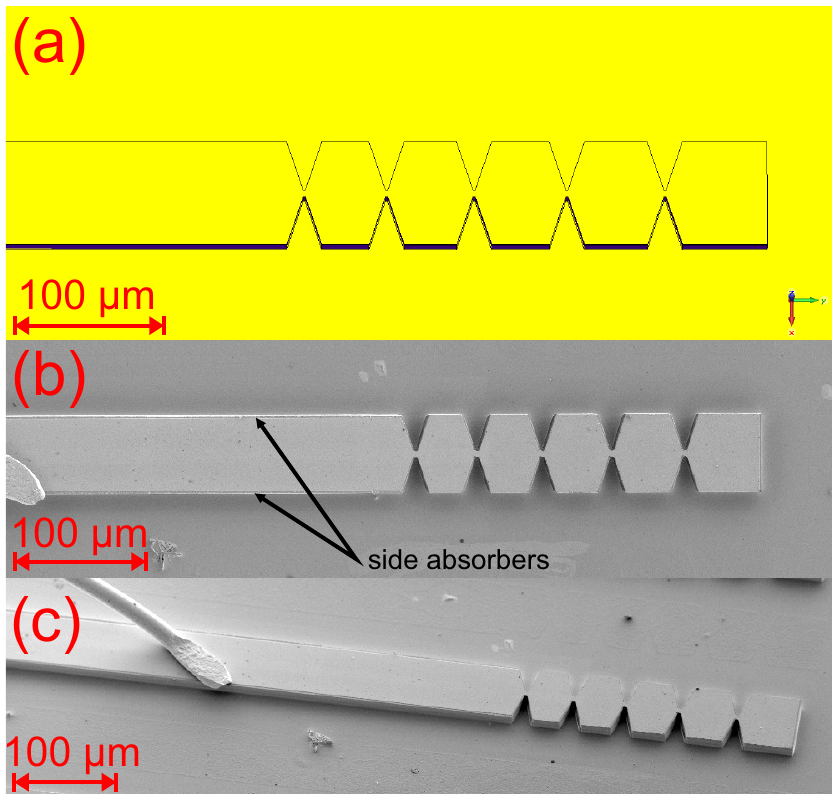}
\caption{Extractor design: (a) Simulation model used in CST Microwave Studio$^{\text{\textregistered}}$. (b)-(c) Scanning electron microscope images of the fabricated structure.}
\label{fig:sim}
\end{figure}
Simulations have been performed using the 3D finite-element solver CST Microwave Studio$^{\text{\textregistered}}$. The simulations have been used to calculate the expected far-field pattern of the designed extractor over the frequency range of interest. Starting from a pure 3rd order Bragg grating, the design has been modified to make it usable for a broad frequency range of 600\,GHz centered around 2.6\,THz. In this range the far-field should be directional in a forward direction, ideally single-lobed with a small convergence angle. \\
The reported, optimized design consists of deep-etched triangles rather than rectangular slits as in the case of patch antennae or DFBs \cite{amanti2009,bonzon2014patch}. This relaxes the 3rd order Bragg condition, making it compatible with broadband frequency operation while keeping a directional beam in a forward direction. Additionally, a longitudinal chirp has been added to the grating, increasing the grating period by 3\,\textmu m every period. The main limitation for the bandwidth is the 2nd order Bragg condition which will cause an extraction shifted by 90\textdegree. The unpatterned section of the 60\,\textmu m wide waveguide features side-absorbers to prevent lasing on higher order lateral modes \cite{bachmann2016setback}. The side-absorbers were introduced by FIB and a deposition of 5\,nm of Nickel (see Figure \ref{fig:sim}(b)). Figure \ref{fig:sim} shows the simulated structure (a), along with scanning electron microscope (SEM) images (b)-(c) of the fabricated extractor. A standard fabrication has been used, including a deposition of a sacrificial silicon nitride (SiNx) etch mask, and a inductivly coupled plasma (ICP) dry etching step.\\  
Figure \ref{fig:farfield}(a) shows the simulated far-field pattern at a frequency of 2.5\,THz for the extractor. A single-lobed far-field with a full-width at half-maximum (FWHM) of 15\textdegree x 40\textdegree \,is predicted by the simulations. To estimate the limits of the extracting structure the farfield was simulated as a function of frequency \cite{footnote}. A forward-directed farfield is achieved only for frequencies between 2.2-2.8\,THz, defining the bandwidth of the extractor. \\
The fabricated lasers with extractor were measured to confirm their simulated properties. Hence, the far-field pattern was recorded. The laser was driven close to its maximal operation point in a quasi-continuous wave (quasi-CW) configuration. A micro-pulse (200\,kHz, 95\% duty cycle) macro-pulse (30\,Hz, 50\% duty cycle) scheme has been used to enable a lockin detection with the used pyroelectric detector (Gentec-EO: THZ2I-BL-BNC). The detector is then scanned around the laser facet with the extractor using two rotational stages. The measured far-field, reported in Figure \ref{fig:farfield}(b), is in good agreement with the simulation in Figure \ref{fig:farfield}(a). A FWHM of approximately 25\textdegree x 30\textdegree \,is achieved with a single lobe. Compared to standard metal-metal waveguide far-fields, the reported far-field is mcuh stronger confined. \\
This large improvement in far-field has also a strong impact on the collectable power, e.g. with a off-axis parabolic mirror ($f/3$), of such a laser. To demonstrate that, the light-current characteristics of the laser were measured while collecting and focusing the light using a off-axis parabolic mirror ($f/3$) each, as displayed in the inset of Figure \ref{fig:liv}. The measurement is then compared to a conventional waveguide laser with similar active area and the same active region using the same setup. These measurements were carried out in continuous wave (CW) operation at a temperature of 15 Kelvin. A calibrated pyroelectric detector (Ophir: 3A-P-THz) was used to measure the absolute power in both measurements. \\
Comparing the laser with extractor to the reference laser shows a five-fold increase in maximal output power. A maximal power of 1.1\,mW in CW is measured. This increase can be expected due to the more directional far-field of the laser with extractor. The difference in the current-voltage characteristics of the two lasers in Figure \ref{fig:liv} arises from slightly different contacts resulting from two independent fabrications of the lasers.\\
Also the spectral performance significantly changes if an extractor is implemented. Figure \ref{fig:spectra}(a) shows an FTIR spectrum in CW at the maximal operation point of the laser. Lasing on a spectral bandwidth of 670\,GHz is observed. This is significantly lower than the spectral bandwidth of standard ridges, which can span more than 1.7\,THz for the same active region \cite{Roesch2014,bachmann2016setback}. Essentially, the spectral bandwidth is limited to the bandwidth for which we expect a directional farfield. Simulations of the facet reflectivity using the same model as for the far-field simualtions show a strongly reduced reflectivity outside this bandwidth compared to standard waveguides (see Figure \ref{fig:spectra}(b)). The reduced reflectivity outside the design range of roughly 500-600\,GHz around 2.6\,THz increases the mirror losses drastically at these frequencies, preventing lasing outside the design range.\\
An important aspect of the presented extractor is its compatibility with FC operation of the laser. To probe the comb performance of the laser with an integrated extractor, the radio frequency (RF) signal of the repetition rate of the laser was measured as a function of the driving current. A single narrow beatnote for a fixed current is a strong indicator of FC operation \cite{Hugi2012,Burghoff2014,Roesch2014,Roesch2016}. As shown in Figure \ref{fig:beatnote}, a single narrow beatnote can be observed for more than 50\% of the dynamic range of the laser, indicating FC operation. For some specific currents more than one narrow beatnote is observed which could hint a multi-comb formation in the laser \cite{li2015combdynamics}. In the range of 380\,mA to 400\,mA no beatnote is observed. For this range of driving currents the optical spectrum shows only modes at a multiple spacing of the actual laser repetition frequency. Similar behaviour has also been observed on lasers without extractors using the same active region \cite{li2015combdynamics}. A similar behaviour has lately been observed, and theoretically described in mid-infrared QCLs \cite{mansuripur2016}. \\
Most importantly, also for high bias currents a narrow beatnote can be observed up to the maximal operation point. Therefore the full optical power is also available for FC operation, making this laser an efficent and powerful source for FC generation at THz frequencies. Simulations of the introduced group delay dispersion (GDD) of the extractor structure confirm a flat dispersion over the observed laser spectral bandwidth\cite{footnote}. The calculated GDD is mostly below $2 \cdot 10^{5}$ fs\textsuperscript{2}. The additionally introduced GDD is smaller than the values of the order of $\sim 2 \cdot 10^{5}$ fs\textsuperscript{2} of the bare laser ridge \cite{Roesch2014,bachmann2016dispersion}. The total dispersion is therefore still low enough to allow FC operation on the limited spectral bandwidth of the laser with extractor.\\
We have presented here a broadband extracting structure for THz QCLs. The reported design is also compatible with FC operation. A comparison to standard metal-metal waveguides reveals a 5-fold increase in output power. This power increase can mainly be attributed to a significantly improved far-field pattern. A single-lobed, almost Gaussian shaped beam profile was measured with a FWHM of 25\textdegree x 30\textdegree. \\
The FC operation of a laser with extractor is increased to more than 50\% of the dynamic range. Also at the maximal operation point comb operation is observed with a spectral bandwidth of 670\,GHz at a center frequency of 2.5\,THz with a total power in CW operation of 1.1\,mW. This makes such a laser the ideal candidate for FC based spectroscopy experiments such as dual-comb spectroscopy at THz frequencies. Further improvement of the extractor design might also help to improve the spectral shape to get a more evenly distributed power over the individual modes and could also be used to introduce additional dispersion to improve the FC performance of such lasers.

\section*{Acknowledgement}
The presented work is partially funded by the EU research project TERACOMB (Call identifier FP7-ICT-2011-C, Project No.296500) and by the Swiss Natinal Science Foundation (SNF) through project 200020 165639. The funding is gratefully acknowledged. The authors would like to thank J. Keller for providing the SEM images.

\end{document}


\title{Broadband monolithic extractor for metal-metal waveguide based terahertz quantum cascade laser frequency combs: Supplementary material}%

\author{Markus R\"osch}%
\email{mroesch@phys.ethz.ch}
\author{Ileana-Cristina Benea-Chelmus}
\author{Christopher Bonzon}
\author{Martin J. S\"uess}
\author{Mattias Beck}
\author{J\'er\^ome Faist}
\author{Giacomo Scalari}
\email{scalari@phys.ethz.ch}
\affiliation{ETH Zurich, Institute of Quantum Electronics, Auguste-Piccard-Hof 1, Zurich 8093, Switzerland}

\maketitle

\section*{Frequency-dependent farfield simulations}
\label{ch:farfields_ex}
\begin{figure*}[tbh]
  \centering
  \includegraphics[scale=1]{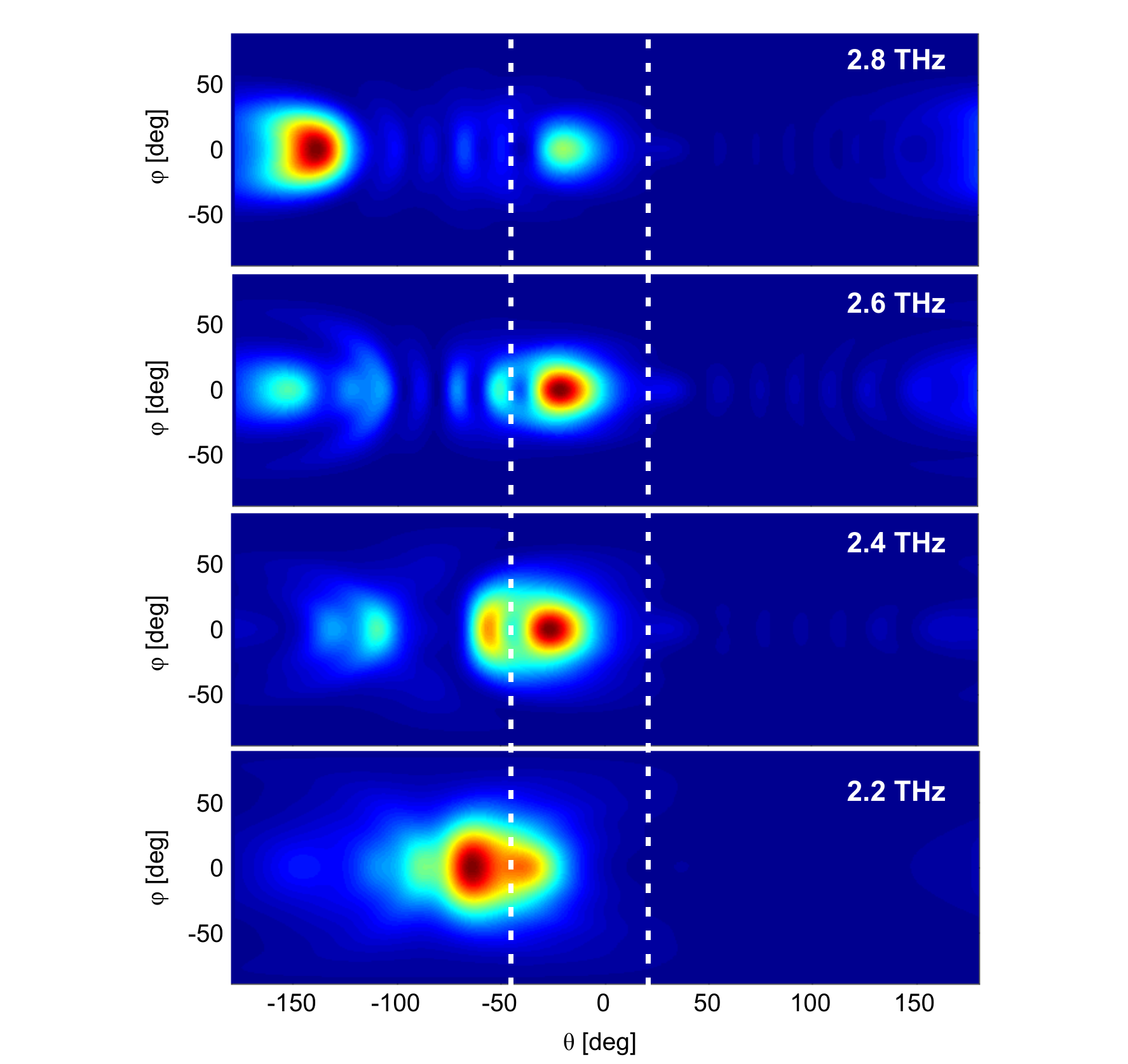}
\caption{Frequency-dependent farfield simulations of the extractor: Full solid angle of the simulated farfield pattern of the extractor simulated with \textsc{CST Microwave studio}$^{\text{\textregistered}}$ for 2.2, 2.4, 2.6, and 2.8\,THz. The dashed lines indicate the range observed in experimental farfield measurements shown in the main text.}
\label{fig:farfield_ex_freq}
\end{figure*}

Here we present frequency-dependent farfield simulations of the reported extractor structure. The simulations are displayed over the full solid angle as a function of frequency in Figure \ref{fig:farfield_ex_freq}. At the lower and upper limit of the frequency range (2.2\,THz and 2.8\,THz) a major part of the light is emitted at a larger angle $\theta$ i.e. surface emission in the case of 2.2\,THz and backward emission in the case of 2.8\,THz. For frequencies inbetween a forward emission at an angle of $\theta \approx 10-20$\textdegree\,is expected.
\section*{Dispersion simulations}
\label{sec:dispersion_extractor}
\begin{figure*}[tbh]
  \centering
  \includegraphics[scale=1]{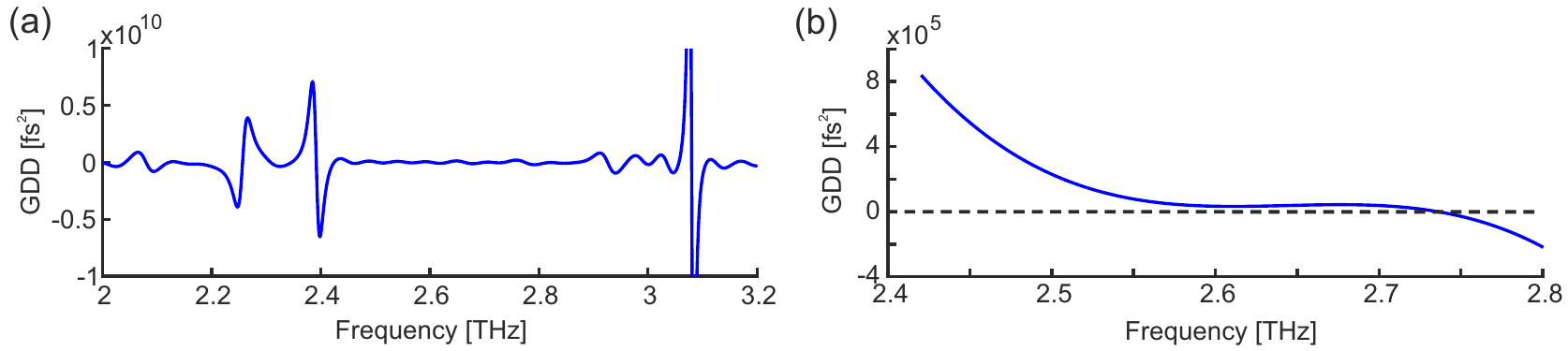}
\caption{Frequency-dependent dispersion simulations of the extractor: (a) The phase of the extractor was simulated with \textsc{CST Microwave studio}$^{\text{\textregistered}}$. From this phase the GDD was calculated. (b) Zoomed version of the data in (a) where the phase was fitted before the derivations with a polynomial function of 5th order.}
\label{fig:disp_ex_freq}
\end{figure*}

To understand better the impact on dispersion by the extractor structure, simulations were carried out to estimate the group delay dispersion (GDD) of the extractor structure. The 3D finite element solver \textsc{CST Microwave studio}$^{\text{\textregistered}}$ was used to calculate the introduced phase $\phi$ by the extractor. The GDD was calculated by two derivations of the simulated phase. The resulting GDD is shown in figure \ref{fig:disp_ex_freq}(a). The strong dispersion peaks below 2.4\,THz and above 3.0\,THz together with the reduced reflectivity at these frequencies are most likely causing the laser only to show modes within these boundaries. \\
A zoomed version of the calculated GDD within these boundaries is shown in figure \ref{fig:disp_ex_freq}(b). To suppress artefacts from the finite step size of the simulation, which are dominating after calculating the second derivative, the phase was fitted with a polynomial fuction of 5th order. The estimated dispersion is positive and of the same order of magnitude as the one of the cold cavity in lasers without extractors \cite{Roesch2014,bachmann2016dispersion}. The added dispersion is small enough to still allow frequency comb operation over the limited bandwidth of the lasers with extractors. For a better understanding of the impact on the dispersion, measurements as in Ref. \cite{bachmann2016dispersion} could be performed on lasers with extractor.